\documentclass[%
reprint,
superscriptaddress,
showpacs,
 amsmath,amssymb,
prl,
]{revtex4-1}

\usepackage{graphicx}
\usepackage{dcolumn}
\usepackage{natbib} 
\usepackage{textcase,bm}
\usepackage{url}
\usepackage[ddmmyyyy]{datetime}
\usepackage{caption}
\usepackage{sidecap}
\usepackage{enumerate}
\usepackage{fancyref}

\renewcommand{\vec}[1]{\mathbf{#1}}

\def\clf{Central Laser Facility, STFC, Rutherford Appleton Laboratory, Chilton, Didcot, OX11 0QX, UK.}
\def\strath{SUPA, University of Strathclyde, Glasgow, Scotland, 4G 0NG,U.K.}
\def\RALSpace{RAL Space, STFC, Rutherford Appleton Laboratory, Chilton, Didcot, OX11 0QX, UK.}
\def\IST{GoLP/Instituto de Plasmas e Fus$\tilde{a}$o Nuclear, Instituto Superior T{\'e}cnico, Universidade de Lisboa, 1049-001 Lisbon, Portugal}
\def\Jasper{Department of Physics and Astronomy, 414 Van Allen Hall, University of Iowa, Iowa City, IA 52242, U.S.A. }
\def\Georgina{The Lunar and Planetary Institute, USRA, 3600 Bay Area Blvd, Houston, Texas 77058, USA}

\def\Lisbon{DCTI/ISCTE - Instituto Universit\text{\'a}rio de Lisboa, 1649-026 Lisbon, Portugal}
\def\StAndrews{University of St Andrews, North Haugh, St.~Andrews,
Fife, KY16 9SS, UK.}

\def\Erika{Department of Earth and Space Science, University of Washington, Seattle, WA 98195-1310, USA}

\preprint{RAL Space}

\begin{document}

\title{Formation of Lunar Swirls}

\author{R.A. Bamford}
        \email{Ruth.Bamford@stfc.ac.uk} 
        \affiliation\RALSpace
\author{E.P. Alves}
        \affiliation\IST
\author{F. Cruz}
        \affiliation\IST        
\author{B. J. Kellett}
	 \affiliation\RALSpace	
\author{R. A. Fonseca}
        \affiliation\Lisbon
\author{L.O Silva}
        \affiliation\IST
\author{R.M.G.M. Trines}
        \affiliation\clf
\author{J.S. Halekas}
        \affiliation\Jasper
\author{G. Kramer}
        \affiliation\Georgina
\author{E. Harnett} 
        \affiliation\Erika        
\author{R.A. Cairns} 
        \affiliation\StAndrews
\author{R. Bingham}
        \altaffiliation[Also at ]\clf
        \affiliation\strath
        
\date{\today}

\begin{abstract}

\section{Abstract}

In this paper we show a plausible mechanism that could lead to the formation of the Dark Lanes in Lunar Swirls, and the electromagnetic shielding of the lunar surface that results in the preservation of the  white colour of the lunar regolith. 
 We present the results of a fully self-consistent 2 and 3 dimensional particle-in-cell simulations of mini-magnetospheres  that form above the lunar surface and show that they are consistent with the formation  of `lunar swirls' such as the archetypal formation Reiner Gamma.  The simulations show how the microphysics of the deflection/shielding of plasma operates from a kinetic-scale cavity, and show that this interaction leads to a footprint with sharp features that could be the mechanism behind the generation of `dark lanes'. The physics of mini-magnetospheres is described and shown to be controlled by space-charge fields arising due to the magnetized electrons and unmagnetized ions. A comparison between model and observation is shown for a number of key plasma parameters.

\subsection{Popular Summary}

On the near side of the Moon, located on one of the dark, most ancient plateaus of the Moon, there is an unusual white, swirl marking that appears to bear no relation to any typical impact craters or lunar ejecta. Visible from Earth, the Reiner Gamma Formation, is the quintessential example of a lunar swirl. 

The lunar surface darkens with age due to space weathering from the bombardment of solar wind protons. The presence of small, 10-1000km, regions of unweathered lunar surface where no topographical cause can be determined, have been an intriguing astronomical phenomena for centuries. 

During the early Apollo missions it was discovered that the location and size of lunar swirls, coincide with small ($\sim$~100s~$km$) surface patches of crustal magnetic field. The Moon is without either an atmosphere or an overarching magnetic field, unlike the Earth's, that would erode and disperse differences in rates of space weathering maturation. The lunar swirl markings therefore represent a time integrated record of the ability of low intensity magnetic fields ($\sim 1/100^{th}$ that at the Earth's surface) to shield parts of the Moon's surface from proton bombardment. The fine detail visible within the swirls  points to a very precise and consistent control of the proton flux. Understanding how this works could lead to the ability to artificially recreate and enhance the protective effects for people and instrumentation susceptible to damage from the more intense parts of the cosmic radiation.

In this paper 2D and 3D particle-in-cell plasma simulations are used to show how finite Larmor orbit plasma effects reproduce the small scale deflection, reflection and retardation of incoming interplanetary plasma with the detail level of the lunar swirls and the plasma observations recorded by surveying spacecraft.

\end{abstract}


\maketitle

\subsection{Introduction}

The Moon does not, and may never have had, an active core dynamo with which to generate its own global magnetic field \cite{lyon1967explorer}. During survey missions for Apollo, however, it was discovered that there were several small, static regions of magnetic field on the surface of the Moon~~\cite{dyal1970apollo, coleman1972}. The dimensions of these regions of magnetic field are of the order of 100s of kilometres. This suggests that they originate from  magnetised material on, or very close to, the lunar surface. These small patches of crustal magnetic field have been identified with the formation of ``miniature magnetosphere" cavities~~\cite{lin1998,wieser2009,wieser2010} and show many observational characteristics of mini-magnetosphere boundaries~~\cite{saito2008,saito2010,futaana2003,ChangeE1,hashimoto2010,kallio2012}. This is despite their relatively small dimensions, many orders of magnitude smaller in dimensions than their planetary cousins, and much lower magnetic field intensity (believed to be only of the order of 100-300nT at the surface). Particular interest in these features is provided by the apparent link between the mini-magnetospheres and the creation of `lunar swirl' patterns on the Moon~~\cite{el_baz1972new,hood1980,hood1989,blewett2011}.

Lunar swirls are optically distinct, white, curvilinear surface features that are found in several locations across the lunar surface~\cite{el_baz1972apollo15}. The features are distinctive due to their fluid or wispy structure that is unlike either craters or impact ejecta. Their form has been determined to be unrelated to geographical topography and appears to overprint on both mountainous and plateau terrain~\cite{bell1987recent,pinet2000local,blewett2007}. What is found associated with all lunar swirls is that they are always co-incident with similar sized areas of magnetic field~~\cite{blewett2007}. However not all the anomalous crustal magnetic fields on the Moon have identifiable lunar swirl discolourations~~\cite{kramer2011}.

\subsection{Lunar swirls and magnetic anomalies}

The presence of small areas of magnetic field on an otherwise unmagnetised planetary body is not unique~\cite{acuna1999Mars,starukhina2004swirls,kivelson1995}. The distribution of magnetic field anomalies on the Moon, vary from irregular conglomerations and clusters to small and isolated - such as the Reiner Gamma formation~\cite{hood1980,richmond2003,hood2008,halekas2001}. The largest distributions of crustal anomalies are located on the southern part of the farside of the Moon, antipodal to the Crisium, Serenitatis, Imbrium, and Orientale basins~~\cite{hood1980,richmond2003,hood2008,halekas2001} . 

Several theories exist to explain the creation of of swirls~\cite{schultz1980cometary,starukhina2004swirls,garrick2011}. However, recent work~~\cite{glotch2015} on data from the Lunar Reconnaissance Orbiter Diviner Lunar Radiometer support the hypothesis~~\cite{blewett2007, kramer2011} that the mechanism of the variations in albedo is related to differential solar wind particle bombardment of the lunar regolith. 

The continuous bombardment by solar wind ions and micrometeorites alters the surfaces of all airless bodies in the solar system~~\cite{pieters2003}. The effect of this space weathering include an overall decrease in visible to near‐IR reflectance, attenuation of mafic absorption features and introduction of a strong positive slope (spectral reddening)~~\cite{pieters1993}. These changes are attributed to the production amorphous coatings on grain surfaces and tiny blebs of metallic iron, known as nanophase iron (npFe0)~~\cite{hapke1973,taylor2001lunar}. The corollary of this process is that a reduced particle flux should lead to lighter coloured (i.e.unweathered) regolith, and an enhanced flux should lead to even darker regolith than the backgound~\cite{blewett2007,richmond2008,kramer2011,glotch2015}. 

The fine details visible within the curvilinear shapes of the swirls is often accentuated by dark lanes that wind though the bright swirls. The transition between the light and dark is often very sharp and the width of the dark lanes can be narrow, less than a kilometre. These dark lanes suggest a consistent enhancement in proton bombardment relative to on-swirl reduction and near-by off-swirl surface representing `normal' lunar regolith~~\cite{blewett2007,richmond2008,kramer2011,glotch2015}.

For this level of detail to have formed and been retained by the swirls, points to a very precise and consistent control of the proton flux. The finesse of the transitions further suggests the process occurs close to the surface. A remote plasma structure would be more likely to be shifted and dispersed by the fluctuations in the solar wind and lunar cycle plus effects of transitions in and out of the regions of the Earth's magnetotail. 

The question now becomes determining the mechanism for the consistent, localised, low altitude redirection of the solar wind protons that could produce the distinct patterns of the swirls. This depends on the fundamental plasma physics that determine the electromagnetic structures arising as a result of the solar wind impacting the small scale magnetic anomalies.

The characteristics of this interaction above the surface have been measured by several in-situ spacecraft~~\cite{lin1998, halekas2007,halekas2014, hashimoto2010, saito2012, wieser2010}. This allows direct comparisons to be made between models and measurements. It is clear from the data that the solar wind ions are reflected and deflected by the magnetic fields as with planetary magnetospheres~~\cite{lin1998}. Conclusive evidence that the crustal magnetic anomalies are also capable of producing collisionless shocks has also recently been demonstrated by the analysis of Halekas and co-workers~~\cite{halekas2014}.

In determining the overall form of a planetary magnetosphere, the ions and electrons of the surrounding plasma environment can be approximated as being locked together with each other and the magnetic field lines. However, the small size of the magnetic `bubble' created by the lunar magnetic anomalies is larger than the scale of the electron dynamics ($<1 km-\sim$20 km), but very much smaller than the ion dynamics scale ($\sim$100s-1,000s~km). This makes lunar magnetic anomalies ideal natural laboratories for studying fundamental collisionless plasma physics phenomena and to validate the small scale kinetic processes in particular, finite ion Larmor radius effects of magnetised, collisionless shocks. The equivalent sizes in magnetic confinement or inertia fusion plasmas would be $\sim1 mm$ and $\sim 1 \mu m$.

In a previous paper~~\cite{bamford2012}, in-situ satellite data, theory and laboratory validation showed that it is an electric field associated with the small scale collisionless shock that is responsible for reflecting, slowing and deflecting the incoming solar wind around mini-magnetospheres. It was shown that the electric field of polarization, caused by the gradient in the magnetic field,  between charge carriers of the solar wind flow, is of prime importance. This polarization field leads to reflection and scattering of the protons and electrons~~\cite{bingham2010}. The counter-streaming of the ions ahead of the barrier is responsible for generating lower-hybrid waves via the modified two-stream instability~~\cite{mcbride1972}, a kinetic plasma process.

In this paper our 2- and 3D simulations support the observational evidence and analytical arguments this results in a self consistent explanation for the creation of `lunar swirls". It will be shown how the 3D particle-in-cell plasma simulation reproduces the key observational characteristics from both above and on the surface, using the simplest of magnetic tomography models - a single magnetic dipole. Thus illustrating the complexity observed comes from the plasma-magnetic field interaction at these scale sizes. In the 2D simulations shown, some observables - such as barrier width - can be seen to be consistent irrespective of changes to magnetic field orientation, size and solar wind conditions. Conversely the variability of other characteristics - such as stand-off distance - can alter significantly with environmental plasma conditions as expected. 
The extended swirl pattern can be envisaged as being composed from a scattered distribution of the elemental components (a subset of which are shown here as examples) with varying extent, orientation, field intensity and overlap.

Other features, such as the most distinctive ``eye" of the Reiner Gamma formation~~\cite{reinergamma}, will be shown to be consistent with a single dipole (in 3D), laid on it's side with it's magnetic axis parallel to the lunar surface, interacting with a downward flowing solar wind plasma in accordance with the theoretical principles outlined in \cite{bamford2012} and computationally illustrated here. A recreation is shown of exaggerated typical spacecraft instrumentation signatures for a transit through a simple mini-magnetosphere.

\section{Method}


The plasma simulation was carried out using a particle-in-cell (PIC) code, called OSIRIS~~\cite{osiris}. In OSIRIS the full set of Maxwell’s equations are solved on a grid using currents and charge densities calculated by weighting discrete particles onto the grid. Each particle is pushed to a new position and momentum via self-consistently calculated fields. The code makes few physics approximations and is ideally suited for studying complex systems with many degrees of freedom such as this one. The cost of this fidelity is the computational expense of the larger number of more complex algorithms interacting with each other than in a simulation of a more simple model system. The reason this is necessary is that the scale size of the mini-magnetosphere structure is very much smaller than the hydro-code approximations allow. No filtering is performed in these simulations. this allows us to resolve all waves in space and time including whistler waves. The code is a time and space domain code, not a spectral code, so the equations are integrated via FFTs. 

Mostly shown here are the 3D simulations with cut-throughs (see graphic in Figure~\ref{FIG:Setup}) to highlight some features. The full 3D simulation is important because plasmas inherently have orthogonal acting forces and flows such as $E=v\times B$, and some aspects are missed in 2D simulations (examples of which are shown in Figures\ref{FIG:different_conditions1} to \ref{FIG:magnetotail}).
When the dimensions of the magnetic anomalies are compared to the characteristic scale length of the magnetic inhomogeneity/magnetic anomaly, $\widetilde{\lambda}_B$, then the value obtained for $\widetilde{\lambda}_B = \left [ (1/B_{sw}). (dB/dh)\right ]^{-1}$ is $\sim 1km$, this is assuming a surface magnetic field of 200nT and the boundary to be at an altitude, $h$, of $20km$. 

What this means is the incoming ions have gyro radii of the order of the structure size. Therefore their motions are not simply following the magnetic field lines, they are unmagnetized. However the electrons gyro radius is sufficiently small, compared to the overall size of the structure, that they are able to follow the changes in magnetic field created by the magnetic anomaly. The difference in behaviour between the electrons and ions sets up a space charge that controls the ions behaviour~~\cite{borisov2003, bingham1993}. To represent the dynamics it is necessary to perform either laboratory simulations~~\cite{bamford2008, bamford2012} or full PIC code simulations, like OSIRIS, as is done here. 

The simulation code operates in normalised plasma units. The independent variable being the plasma density $n_{sw}$. The units for distance are normalised multiples of the electron skin depth, $c/\omega_{pe}$ (where $c$ is the speed of light and $\omega_{pe}$ is the electron plasma frequency $\omega_{pe}=\sqrtsign{n_ee^2/m_e\epsilon_o}$ and  $e$ the electric charge, $m_e$ the electron mass and $\epsilon_o$ the permittivity of free space). From which it can be seen that the physical dimensions of the box scale inversely with the square root of the free parameter of plasma density, $n_{sw}$. The equivalent dimensions in kilometres recreated by the simulation can be made to contract with higher densities and expand with lower as other de-normalised parameters, such as magnetic field intensity, would also increase and decrease to remain in proportion with the defined important plasma parameters (specifically the magnetosonic mach number and plasma $\beta $) that determine the correct realm of plasma environment on the Moon.

\subsection{Simulation parameters}

\begin{figure*}[h]
        \includegraphics[width=0.65\textwidth]{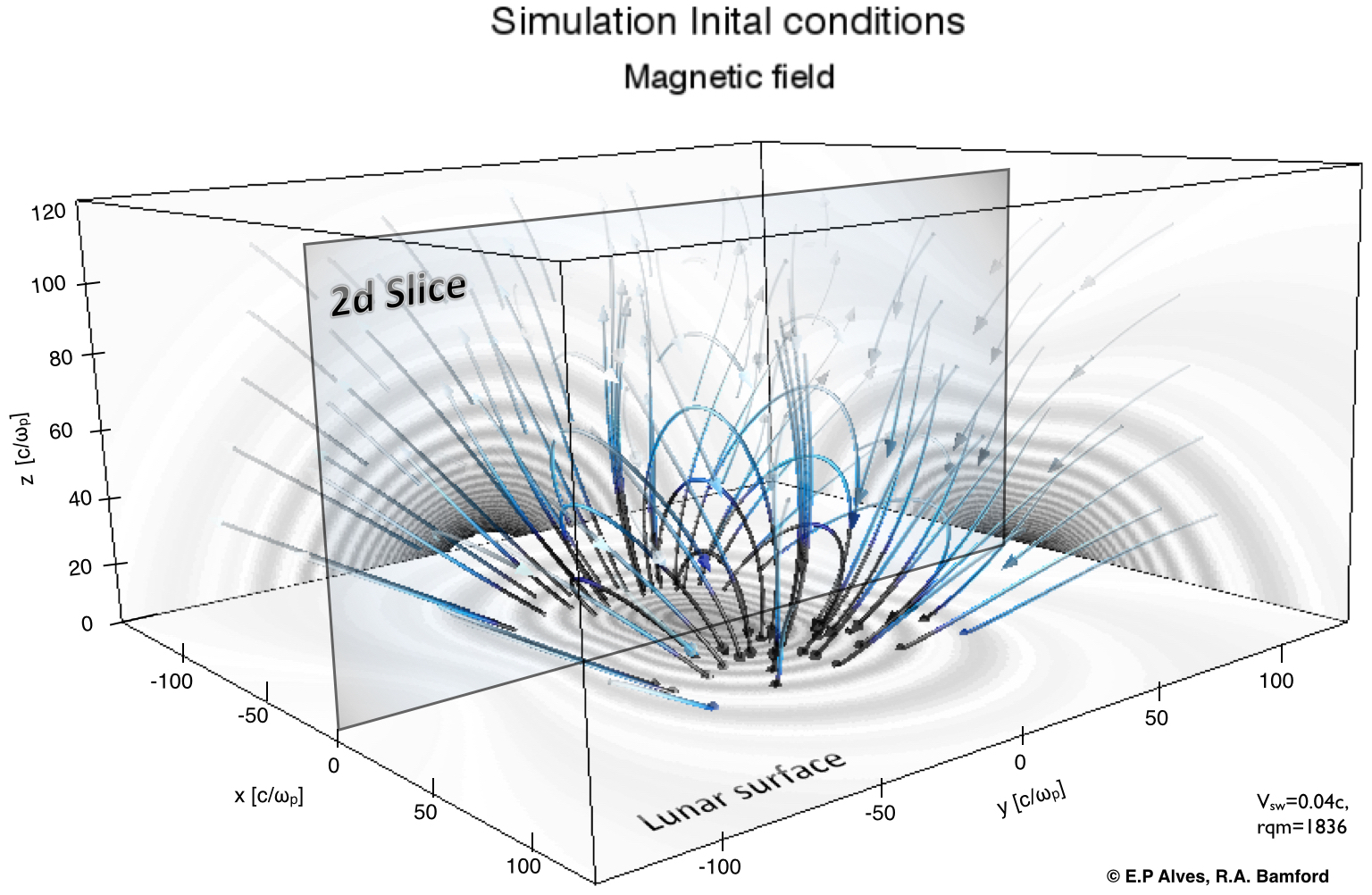}
       \captionsetup{justification=raggedright, singlelinecheck=false}

        \caption{The initial conditions of the simulation. The magnetic field intensity is shown projected onto the back walls and ground plane in banded grey where decreasing band interval corresponds to increasing magnetic field intensity. Selected magnetic field lines are shown in graduated blue. The magnetic dipole moment $\vec{m_m}$ is 25 normalized units long and resides 25 units below the center of the box aligned with the positive $\vec{x}$ axis. The plane marked `2D slice' shows the relative orientation of the sections shown in Figures~\ref{FIG:2dSlice3D} and~\ref{FIG:DarkLanes}. 
}\label{FIG:Setup}      
\end{figure*}

The 3D simulation geometry and initial conditions are shown in Figure~\ref{FIG:Setup}. The lunar surface is represented by the lower $x-z $ plane. A single source magnetic dipole is placed just below and parallel to the $x-z $ plane, with magnetic axis aligned along $x =0$, with the north pole orientated in the $+z $ direction. The result is a hemispherical magnetic field emerging from the surface. A magnetised `solar wind' plasma with density $n_{sw}$ and magnetic field $B_{sw}$ (antiparallel to anomaly field) is introduced from the top plane with a flow velocity $-\vec{v_{sw}}$ vertically down onto the lunar surface. 

Observationally the mean value of the solar wind magnetosonic Mach number is $M_m=8$~\cite{edberg2010magnetosonic}, with a $\beta=0.2$, $n_{sw}=10$cm$^{-3}$, $B_{sw}=10$nT and $T_i=5eV$. The electron and ion skin depths are $=1.7$km and $72$km respectively. The ion Larmor orbits are $97km$ for the thermal distribution and $627$km for the flow velocity. 

This makes $c/\omega_{pe} =1.7$~km and the simulation box shown represents $\sim 340 \times 340$~km by $\sim 170$~km altitude. The grid cell size is 5 cells electron skin depth, i.e., grid resolution of $\sim$300~m. In this simulation the magnetic moment of buried dipole $= 9.10^{15} $A m${^2}$ at a depth of $\sim$50~km (for density of $n_{sw}=10$~cm$^{-3}$). 

The tradeoff for preforming full-pic in 3D required operating within the simulation the plasma flow speed is increased by a factor $F=v_{sw}/v_{osiris}$. In the 3D simulations $F=20$ assuming a $v_{sw}=600$~km~s$^{-1}$. 

In order to maintain dynamic similarity with the lunar environment, the magnetic fields and temperatures were scaled proportionately so as to maintain the same control variables of plasma $\beta$ (thermal pressure to magnetic pressure) and $M_m$ (magnetosonic mach numbers and speeds). 
The fast magnetosonic mach number $M_m=v_{sw}/\sqrtsign{v_A^2 + v_s^2}$ where $v_s$ is the sound speed and $v_A$ is the Alfv{\'e}n speed. For the same plasma density $n$, the temperature within the simulation was $T_{osiris}=T_{sw}F^2$ and magnetic field intensities was $B_{osiris}=B_{sw}F$. The plasma frequency $\omega_{pe}=\sqrtsign{ne^2/m\epsilon_o}$ is unaffected. The real proton-to-electron mass ratio of 1836 was used.

For this simulation the peak magnetic field at the surface, $B_{sf}$, scales as $\sim 2000$nT for a realistic quiet sun solar wind of $v_{sw}=600~$kms$^{-1}$ and $n_{sw}=10$cm$^{-3}$.

The standoff position $r_s$, in the 3D simulation, of the cavity was engineered to be at the ion skin depth $c/\omega_{pi}$ (for a plasma density of $10cm^3$, $r_s=100$km, $B_{surface}\approx2\mu$T). The motivation for this  was that observation of the magnetic compression would occur at the same altitude and completely within the simulation box. 

The consequences of the scaled up velocity are that the simulation surface magnetic field had to be similarly scaled up to match the flow pressure. Also not all the microphysics scales, in particular excited waves, collisionless dissipation mechanisms may be reproduced. But the microphysics of the collective space-charge electric field, which is responsible for the plasma shielding/deflection is well captured. This then results in a sharp foot print with features on $\sim c/\omega_{pe}$ scales as will be shown later.

The analytical expression for $r_s$ related to the pressure balance between incoming solar wind $n_{sw}m v_{sw}^2 = B^2/2\mu_o$, is well established for planetary e.g.~~\cite{bingham1993}, and mini-magnetospheres~~\cite{gargate2008}. Such that: 
\begin{equation}
    r_s \simeq \left ( \frac{ B_o^2}{2n_{sw} m v_{sw}^2} \right )^{1/6}.
\end{equation}\label{EQU:r_s}
Here $B_o$ is the subselenean dipole moment.

Using realistic magnetic field strengths will lead to smaller cavity structures and lower $r_s$. 

\section{Simulations results}

The simulation code is used here to recreate simplified case study combinations of plasma parameter conditions and magnetic field dipole orientations and intensities. These can be compared to the analytical expressions \cite{bamford2012} and observational data e.g.\cite{halekas2014}. Included here are the importance of simulating 3 dimensions in a plasma due to the inherent cross-product relationship of forces. 

\begin{figure*}[h]
     \includegraphics[width=0.75\textwidth]{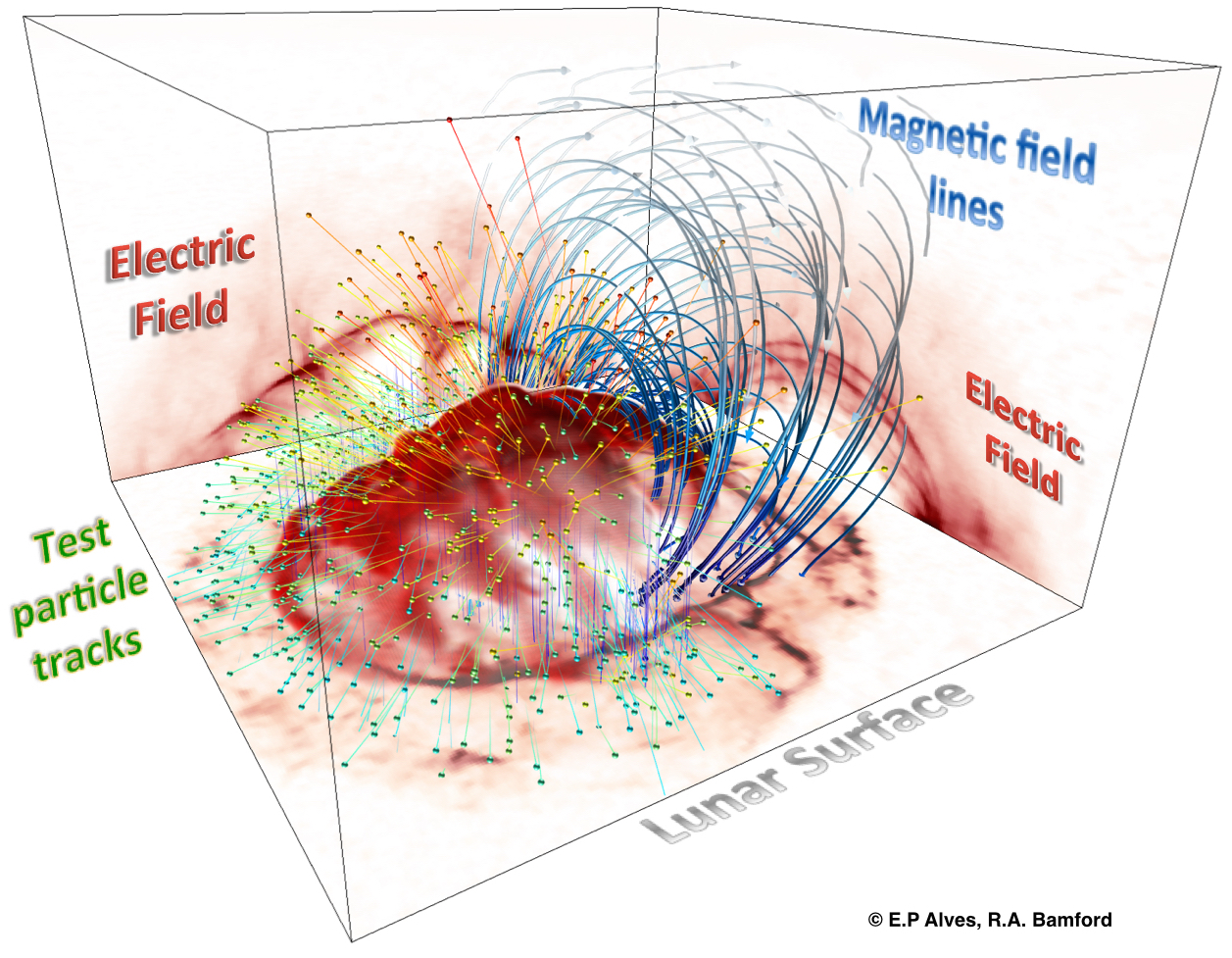}
     \captionsetup{justification=raggedright, singlelinecheck=false}
     
     \caption{(Colour) A 3D magnetized plasma collision with a surface magnetic dipole. The solar wind plasma (flowing vertically downwards) impacts a localised crustal magnetic field structure (blue lines). The green spheres and tracks show a subset of protons population trajectory   being scattered from the narrow polarisation electric field (red). Proton density enhancements follow the electric field. Only part of the magnetic field lines are shown for clarity and the background densities are not visualised.
     }\label{FIG:3D}
\end{figure*}   

Figure~\ref{FIG:3D} shows the results of the simulation of the solar wind plasma impacting a localised crustal magnetic field structure. In Figure~\ref{FIG:3D} only the ion density above a threshold is visualised in order to make the box transparent. The magnetic field structure is shown by the blue field lines (again in-part omitted for sake of clarity). The red represents the space-charge electric field at the boundary, which is set up by the different penetration depth between ions and electrons at the edge of the magnetosphere. The lateral-projections of the electric field structure (corresponding also to the relative proton density) reveal interesting dynamic features and orthogonal asymmetries. The projection on the $y-z$ plane shows a rippled surface structure, due to the diamagnetic electron-ion drift instability which occurs perpendicular to the magnetic field lines~\cite{fabioEPS}. In contrast, the projection on the $x-z$ plane shows a smooth surface structure since the relative electron-ion drift is absent. This illustrates the anisotropic preferences of particular plasma instabilities. However the narrow width of the barrier remains a consistent feature, although not necessarily a single, smooth boundary due to waves, turbulence and instabilities - the magnitude of which alter with specific conditions.

Representative trajectories of a few solar wind protons are shown by the small spheres and yellow track lines and are seen to be widely scattered not by a gradual redirection but ballistically scattered from a very narrow region close to the surface. The electric field responsible for scattering can be seen to be omni-directionally pointing outwards regardless of the magnetic field orientation. This is because it relates to the {\it gradient} in the magnetic field intensity $\nabla{|B|^2} $ and not $\vec{B}$ as is predicted theoretically~\cite{bamford2012}. The projection onto the $x-z$ or surface plane, shows the electric field intensity at the lunar surface. The proton density is controlled, on these scales, by the electric field rather than the magnetic field because the ions are unmagnetised. The simulation here, therefore, shows the distribution of sharp regions of enhanced proton-flux and regions of depleted proton flux.


\begin{figure*}[h]
        \includegraphics[width=0.5\textwidth]{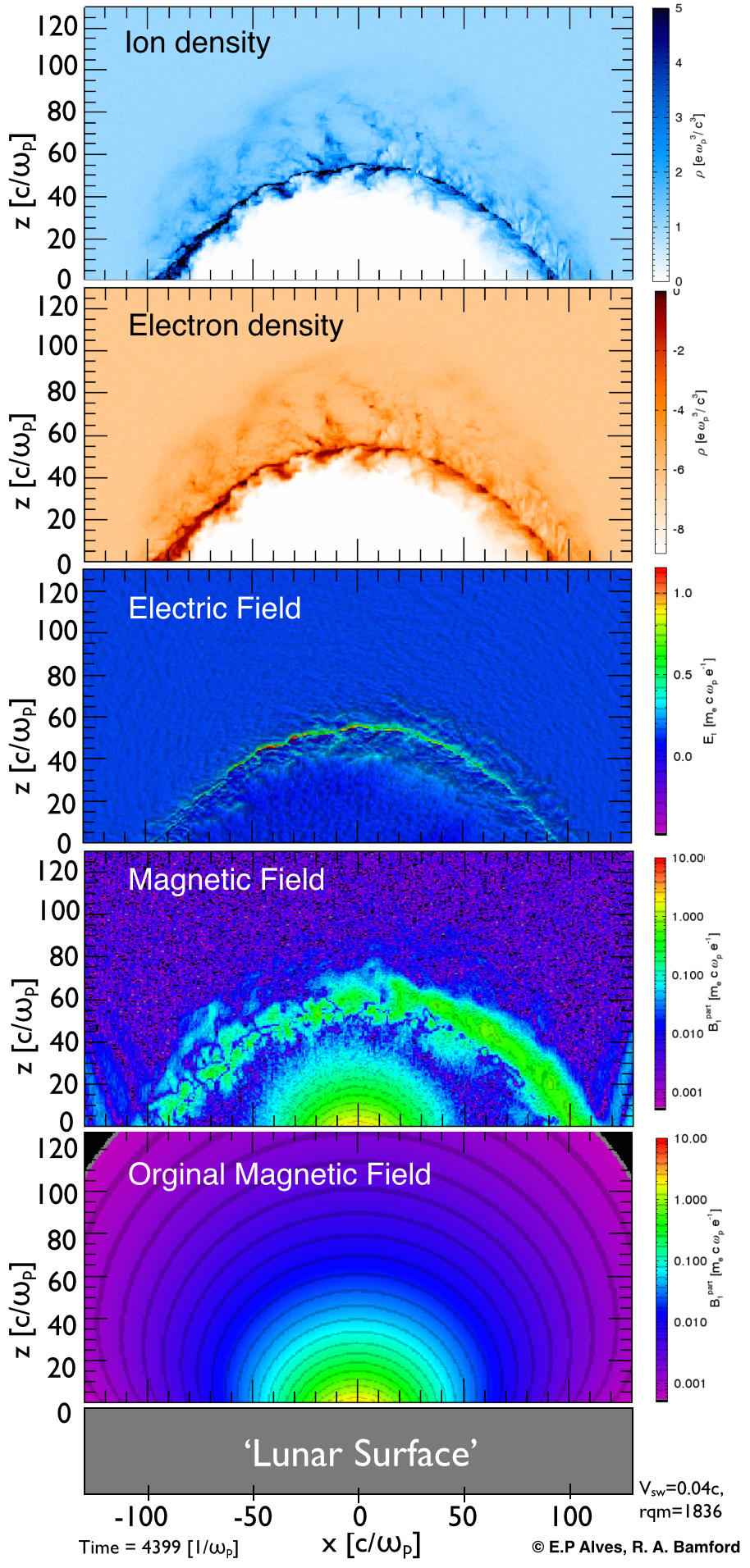}
        \captionsetup{justification=raggedright, singlelinecheck=false}
        
\caption{(Colour) Planar slices through the 3D simulation shown in Figure~\ref{FIG:3D} showing the behaviour of the plasma densities and electromagnetic forces. The location of the plane is indicated in Figure~\ref{FIG:Setup}. The normalised y axis represents altitude above the Moon’s surface. The normalised x axis represents distance along the surface of the Moon upon which a magnetic dipole field is located.
    } \label{FIG:2dSlice3D}
\end{figure*}

In order to reveal the details of the interior structures of the 3D simulation Figure~\ref{FIG:2dSlice3D} shows the 2 dimensional sections of each of the plasma parameters. These are from top to bottom the ion and electron density $n_i$, $n_e$, the resultant electric $E$ and magnetic field $B$ and the original source half dipole magnetic field undisturbed by the solar wind plasma $B_{vac}$. The plane of the 2D section bisects the midpoint in $x$ of the dipole axis and is indicated in Figure~\ref{FIG:Setup} .  

The density pile-ups, exclusion of the majority of the particles from the interior, back-flow and turbulence of the barrier are apparent on all parameters. The barrier currents that form result in further responsive inductive currents and corresponding inductive magnetic fields. The thickness of the barrier is of the order of the electron skin depth $\sim c /\omega_{pe}$, as theoretically predicted~~\cite{bamford2012}. The small-scale plasma instabilities, waves and turbulence formed, provide the means by which the ion and electron particle distributions become non-thermal and exchange energy through Landau damping \cite{bingham1993}. (This is illustrated in the top two panels of Figure~\ref{Fig:flyover} discussed in Section~\ref{SEC:flyover}).

The ``stand-off'' distance, $r_s$ (equation~\ref{EQU:r_s}) that the magnetic cavity reaches force balance with the incoming plasma, is approximately $ 50 c /\omega_{pe}$ by design so as to ensure the interface was well situated within the simulation box.

 
\begin{figure*}[h]
     \includegraphics[width=0.75\textwidth]{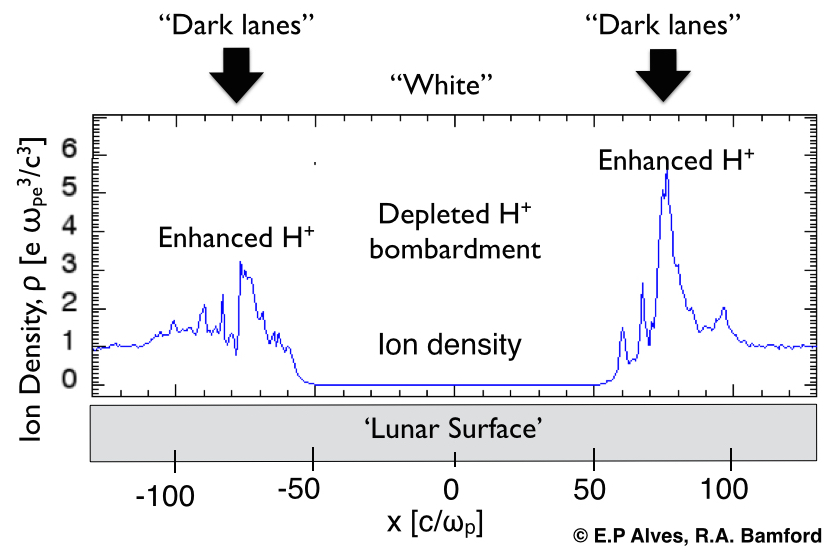}
     \captionsetup{justification=raggedright, singlelinecheck=false}

     \caption{The simulated relative proton deposition onto the surface of the Moon. From the same simulation shown in Figures~\ref{FIG:Setup},\ref{FIG:3D},\ref{FIG:2dSlice3D}.}\label{FIG:DarkLanes}
\end{figure*} 
 
The relative proton density from the simulation can be further clarified by examining a linear plot taken at the equivalent lunar surface level. This is shown in Figure~\ref{FIG:DarkLanes}.

If sustained long term, this pattern of excluded and narrow enhance proton density resulting from the action of the mini-magnetosphere with crustal magnetic field would be consistent with the distribution and finesse required to form the variety of lighter and darker albedo alterations seen in lunar swirls. 
 

\begin{figure*}
     \includegraphics[width=0.75\textwidth]{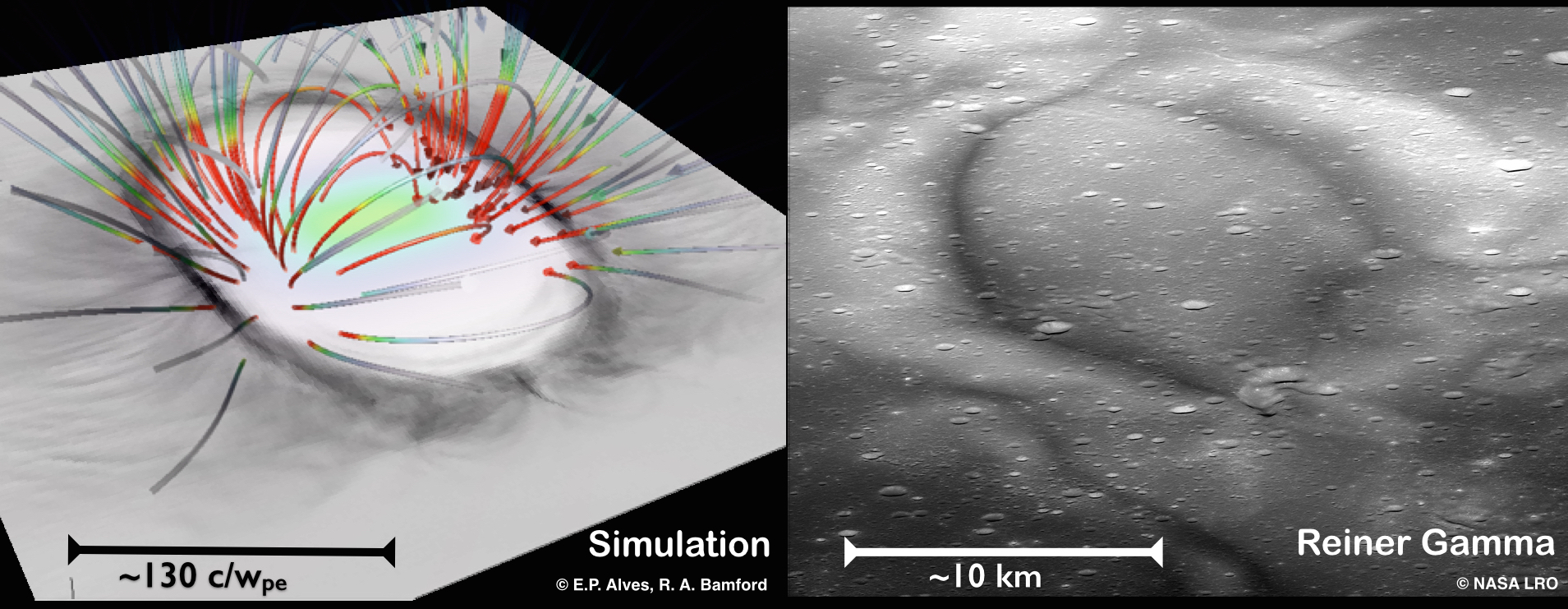}
     \captionsetup{justification=raggedright, singlelinecheck=false}
        \caption{Left:A 2D slice of the relative proton density from the 3D simulation with the initial magnetic field lines from a single subsurface dipole. The greyscale distribution shows darker for higher density of protons, whiter for less. Right: A image of the central region of the Reiner Gamma Formation lunar swirl taken by NASA's Lunar Reconnaissance Orbiter~~\cite{LROC}.The form and relative width of the 'dark lanes'~~\cite{bell1982reiner} suggest the aspect ratio of dark-lane width to cavity width is similar in both cases. 
        }\label{FIG:Swirl_Overlay}
\end{figure*}

The relative deposition of proton flux (shown in grey) on the surface slice from the simulation is shown in the top left image of Figure~\ref{FIG:Swirl_Overlay} with the magnetic field lines of the dipole. This is compared to a image of the central region of the most distinctive example of a lunar swirl, Reiner Gamma Formation located at $7.4^{\circ}N, 300.9^{\circ}E$, taken  by NASA's Lunar Reconnaissance Orbiter~~\cite{LROC}. The agreement between the key characteristics of dark lane~~\cite{bell1982reiner} width and shielded interior can be seen to be totally consistent. The symmetries in the simulation results prevent distinguishing categorically the orientation of the magnetic fields. 

We will now preform a series of simulations to study the parameter regime of single dipoles by varying the orientation and some of the plasma conditions.

\subsection{The effect of dipole orientation}

\begin{figure*}
        \includegraphics[width=1.0\textwidth]{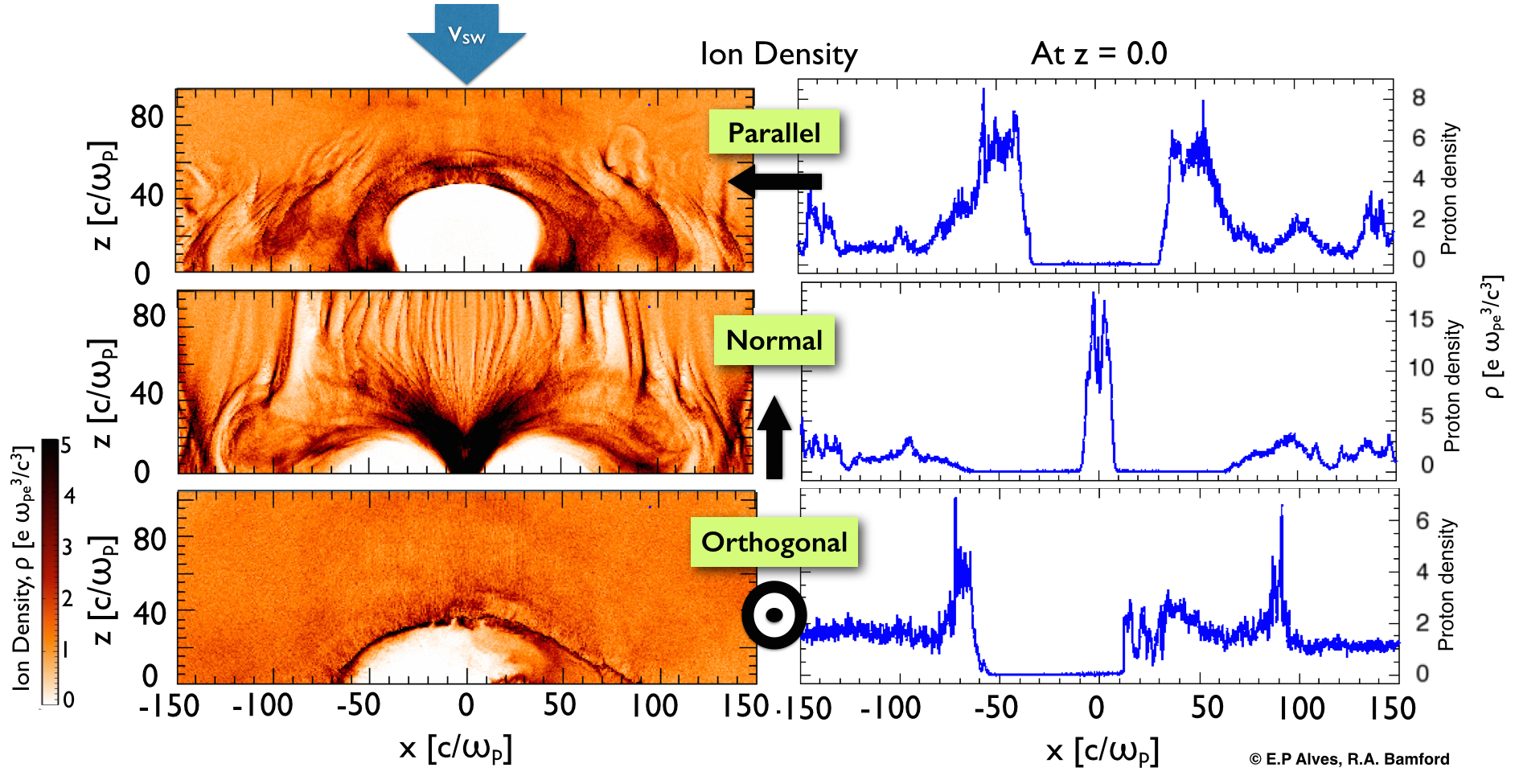}
        \caption{ \raggedright Different orientations of magnetic dipole axis aligned in the x,y,z direction with their corresponding relative proton density ``footprint” on the lunar surface. The double high peak visible in the vertical dipole orientation (middle panel) would in 3d, be the equivalent of the particle deposition pattern that would form an auroral oval.
}\label{FIG:different_orientations}      
\end{figure*} 

Changing the orientation of the emerging magnetic dipole relative to the surface plane produces the same diamagnetic characteristics of narrow barrier, particle reflection, cavity formation, waves and turbulence. However, as can be seen in Figure~\ref{FIG:different_orientations}, the overall morphologies of the relative proton distribution are very different. 

Figure~\ref{FIG:different_orientations} shows 2D simulations of the same magnetic dipole in 3 orthogonal directions $\vec{x}, \vec{y}$ and $\vec{z}$, relative to the surface plane. The solar wind, mass ratio and plasma conditions are as those of the 3D simulation with the exception that the solar wind plasma is not magnetised. The consequence of $\vec{B_{sw}}$ parallel or antiparallel to the surface magnetic field is shown in the next section. 

The linear plots of relative proton density (on the right of the Figure) show different proton deposition profiles for each orientation.

\subsection{The affects of changes in the environmental plasma} 

\begin{figure*}[h]
        \includegraphics[width=0.85\textwidth]{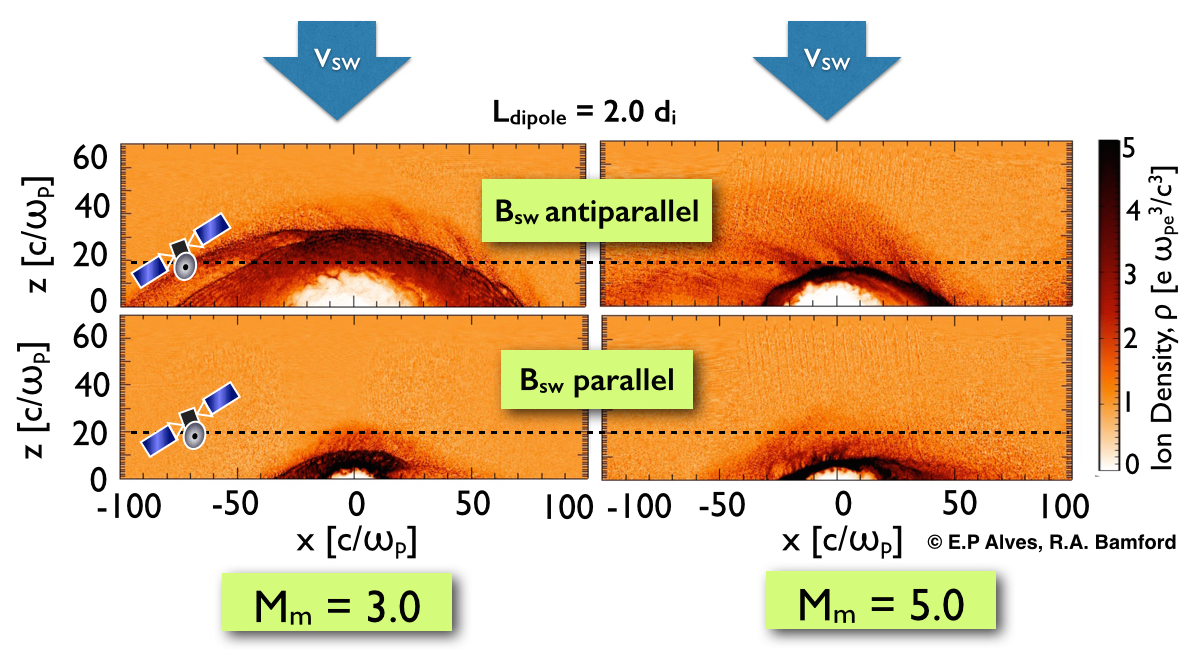}
        \caption{ \raggedright The affect of changes in parameters 1. Perpendicular shocks by cold plasma with fixed dipole size/strength ($L_{dipole}=2$), varying magnetosonic Mach number, $M_m =3, 5$, for parallel and anti-parallel $B_{dipole}$. Reduced 2d parameters: The incoming flow velocity is $\times 50$ a typical solar wind speed, proton to electron mass ratio $m_p/m_e$ is $\sim 1/20$ realistic value, chosen for computational speed. Although these values are not representative of actual values in space, they do allow a qualitative comparison of the variation of mini-magnetosphere characteristics.         
}\label{FIG:different_conditions1}      
\end{figure*}

\begin{figure*}[h]
        \includegraphics[width=0.75\textwidth]{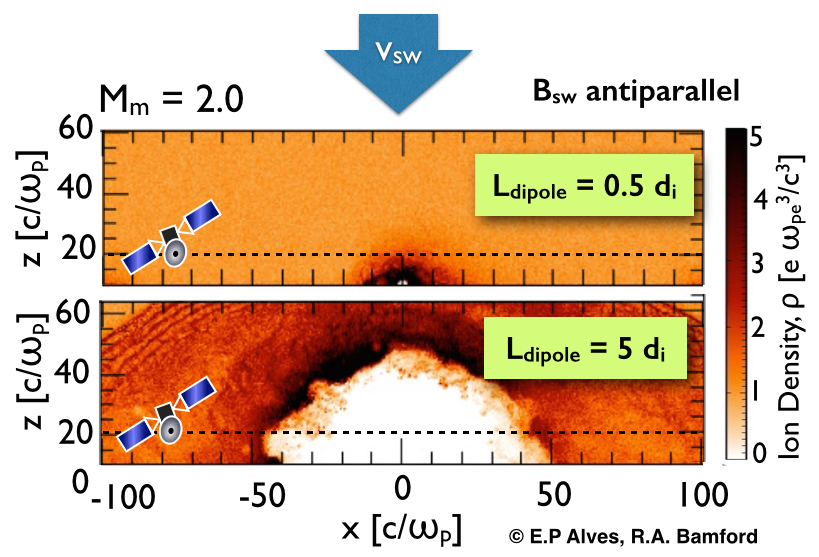}
        \caption{ \raggedright The affect of changes in parameters 2. Perpendicular shocks by cold plasma with fixed magnetosonic Mach number $M_m=2.0$ with $B_{in}$ parallel to $B_{dipole}$, varying dipole size/strength $L_{dipole}=0.5, 2.5, 5.$. Reduced 2d parameters as .         
}\label{FIG:different_conditions2}      
\end{figure*}

\begin{figure*}[h]
        \includegraphics[width=0.75\textwidth]{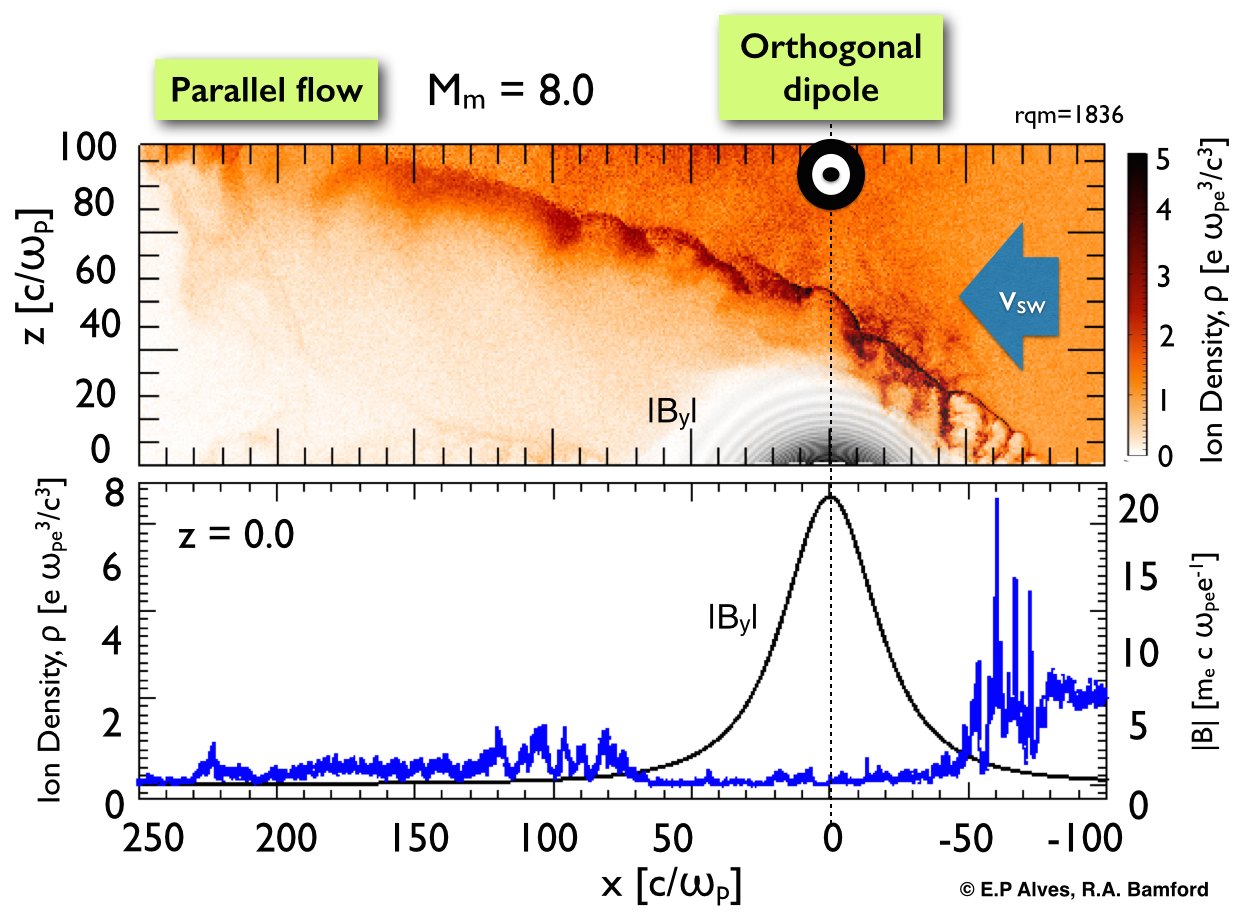}
        \caption{ \raggedright The formation of a mini-magnetotail. The solar wind flow direction is parallel to the lunar surface and the dipole vector is pointing outwards from the page. In the upper panel the magnetic field intensity is shown as banded grey.           
}\label{FIG:magnetotail}      
\end{figure*}

Although the crustal magnetic anomalies are a fixed magnetic field source, the plasma environment is not. There are periodicities due to orbits and diverse solar wind and/or magnetospheric conditions. Through simulation this can be explored, albeit briefly, in Figures~\ref{FIG:different_conditions1}, \ref{FIG:different_conditions2} and~~\ref{FIG:magnetotail}. 

A spacecraft passing over the vicinity of these anomalies, at a fixed height (as indicated in the Figures by a satellite graphic and dashed `flight-path' line) but under the different plasma conditions will sometimes transect the different regions of a mini-magnetosphere and so observe different characteristics. (An illustration of these plasma data instrumentation signatures in such a flyover is shown in Section~\ref{SEC:flyover}).

Figures~\ref{FIG:different_conditions1} and \ref{FIG:different_conditions2} illustrate the consequence of combinations of changes the simulation conditions. In Figure~\ref{FIG:different_conditions1}, the incoming solar wind Magnetosonic mach numbers $M_m$ and  parallel (or anti-parallel) solar wind magnetic field $\vec{B_{sw}}$ orientation are varied. In Figure~\ref{FIG:different_conditions2} the dipole size/length, $L_{dipole}$, is altered. The extent of the dipole can be altered either through being a larger single crustal magnetic field or by multiple smaller conglomerations (such as on the far-side of the Moon) that will appear as a single dipole when observed in sufficient altitude to be in the far-field. 

In all these figures the white-to-red colour distribution represents relative ion density and all the conditions are the same except for those stated as otherwise. 

The majority of the simulation results shown have been with the solar wind plasma impacting normal to the plane of the lunar surface andemerging magnetic field. This is purely for simplicity and consistency so as to example the plasma interaction. In Figure~\ref{FIG:magnetotail} the impacting plasma flow is parallel to the lunar surface and so can be seen to create a mini-magnetotail. From Figure~\ref{FIG:magnetotail} it can be seen that many of the features of the normal instance impacts are also seen in the parallel impact case. However the diversity of regimes and features that can be discussed is greatly increased and more suitable for a dedicated study. This is also true of combinations of magnetic field anomalies and conditions. 

In summary, the 2D simulations show how only occasionally might certain features, like the diamagnetic cavity, be clearly detected in in-situ instrumentation, such as particle detectors and magnetometers. This is especially for a rapid spacecraft transit through very small isolated magnetic field anomalies. However, other instrumentations such as imagers\cite{wieser2010}, that look down onto the features could still detect the characteristic depletion in proton reflection.

\subsection{Simulating Spacecraft flyover signatures}\label{SEC:flyover}

\begin{figure*}
         \includegraphics[width=0.60\textwidth]{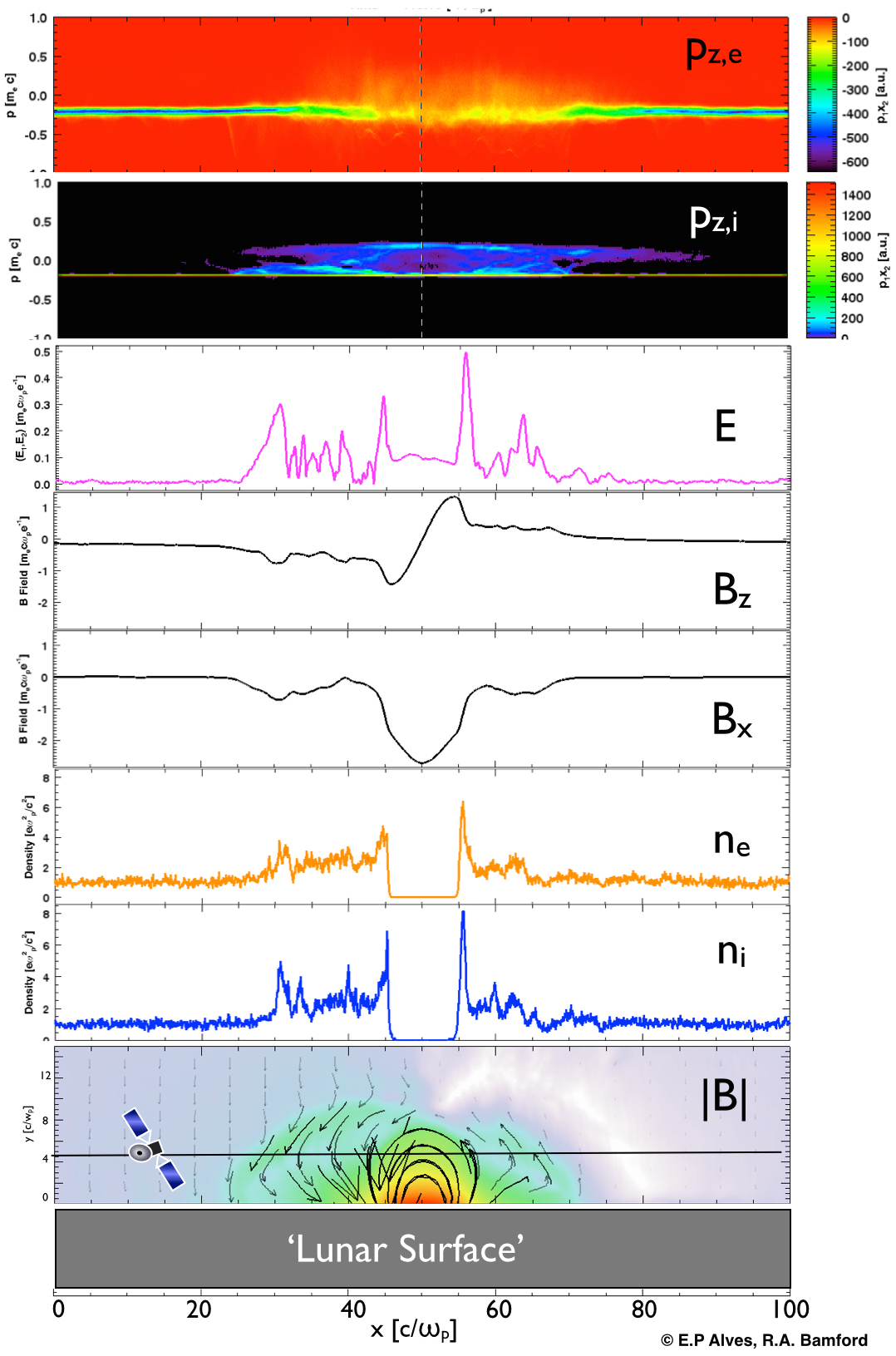}                      \caption{\raggedright (Colour) The simulated spacecraft diagnostic signatures of a transit through a mini-magnetosphere. The 2D OSIRIS simulation is plotted so as to recreate the type of signatures that would be observed by spacecraft plasma instrumentation during a flyover of a crustal magnetic anomaly (lower most panel). A low altitude lunar spacecraft would record in a flyover transit over the surface anomaly at constant altitude of $h = 5.5 \times c/\omega_{pe}$ in normalised units the equivalent of $\sim$12km (for a 5cm$^{-3}$ density plasma). The simulation is the simplest geometry with slow wind flow normal to the lunar surface, there is no drawn out magnetotail in this example. The simulated data window would be the equivalent of $\sim$2 to 4 minutes in duration. }\label{Fig:flyover}  
\end{figure*}

Figure~\ref{Fig:flyover} shows how the simulation results translate as observations to a spacecraft flying above the surface anomaly. 

These features should be common to any mini-magnetosphere to some degree or another irrespective of the presence of lunar swirls. 
 The comparison is qualitative not quantitive to highlight the nature of interplay in the parameters so as to provide identification of phenomena between theory and observation via simulation visualisation.

The plasma signatures reproduced in the simulation are: 
(a) An increase and density pile-up before the entry into the cavity (reported observationally by~~\cite{lin98, saito2012, wieser2010,halekas2014}). This is most evident in Figure~\ref{Fig:flyover} panels labeled ion ($n_i$) and electron ($n_e$) densities. 
(b) The magnetic field components (panels labelled $B_z$ and $B_x$ in Figure~\ref{Fig:flyover}) are seen to rotate as if ``draped'' around a small magnetic obstacle as first reported by Lunar Prospector~\cite{lin98}. (c) An increase in electrostatic ($E$) field and solitary waves at lower hybrid frequencies appear (observational data~\cite{lin98,halekas2008,wieser2010,futaana2003}). 
(d) Non-adiabatic energy between the ions and electrons \cite{saito2012, halekas2014}. In the upper most panel, the electron momentum $p_{z,e}$ in the vertical/flow ($z$) direction, increases and the energy distribution (temperature) changes as seen by the spreading in $p_{z,e}$. 
(e) Back streaming protons \cite{saito2012, halekas2014}   accelerated by similar factors close to the shock surface~~\cite{wieser2010}. The protons momentum $p_{z,i}$ (next panel) shows how the ions are reflected back from the magnetic structure predominantly at incoming velocity. 
(f) Electrostatic oscillation at the lower hybrid frequency (seen observationally \cite{saito2012}). The back-flow of ions establishes the conditions for the modified two stream instability that oscillates at the lower hybrid frequency~~\cite{mcbride1972}. 
 
The simulation operated with a $v_{sw}$ $\times 100$ the typical realistic quite time velocity of $600$~kms$^{-1}$ and a reduced mass of 100. This emphasised in the simulation, the non-adiabatic energy between the ions and electrons (top two panels).

\section{Conclusions}

We have reproduced all the major the characteristics of lunar swirls using the simplest of magnetic topologies - a single dipole.

The simulations confirm many of the satellite findings and the theoretical predictions [5] that a collisionless shock forms at the altitude expected from the theory of collisionless shocks. The thickness of this shock is approximately equal to the electron skin depth  $c/\omega_{pe}$ (where $\omega_{pe}$ is the electron plasma frequency) again in agreement with theory. The characteristic observations of electron and ion density enhancements and depletions accompanied by magnetic field intensity pile-up at the shock coincide with the formation of a narrow interface region where a dynamically stable electric field exists. The simulations confirm that it is this electric field that controls the behaviour of the solar wind ions impacting the magnetic structure. 

Because the primary driving term is a gradient in energy density, it is not exclusive to one magnetic field orientation. There are asymmetries in the diamagnetic cavity due to a difference in the preferred plasma instabilities and different growth and instability interchange rates. 
 
The 3D simulation has the solar wind flowing from directly above the anomaly, with normal incidence to the surface. The interaction naturally leads to an asymmetric structure in the fields and density because the plasma electrons are preferentially deflected to one side of the magnetic field rather than the other. Though the magnetic structure is symmetric, the plasma response is intrinsically asymmetric. This does not include a magnetotail from an oblique impact, however it is the most basic prototype example.

In the case of the Moon, different orientations of the magnetic field structure to the (a) surface of the Moon and (b) to the incident solar wind - will result in more complex and filamentary structures, particularly near magnetic poles. The size of the filamentary structures could be less than 1 km or even 100 m. This allows a complex surface magnetic field to provide variable levels of protection from the ageing process of the solar wind. What is seen on the surface is the legacy of structures above and the periodicities and variations encountered over time.

A prediction from this work is that the dark lanes should be always approximately the same width ($\sim c/\omega_{pe}$) everywhere on the Moon. Whereas the extent of the  white regions will depend upon magnetic field strengths and near field topography. Future work is to determine the near-field tomography, ideally with in-situ measurements on the ground. We will also look at more complex magnetic configurations and combinations plus mini-magnetotail formation in a future papers.

Although not all the microphysical phenomena (e.g. dissipation, excited waves, etc) scales correctly in the simulation nor rely on these simulations to extract realistic particle distribution data, and exact growth rates of the excited waves upstream of the reflection of the plasma, etc.  Although these are first principle simulations, they are still unable to capture all the physics quantifiably exact, because it is still currently impossible to simulate the realistic physical parameters due to the massive computational requirements. 

These simulations do illustrate clearly, however, the microphysics underlying the shielding/deflection of the plasma around a kinetic-scale cavity.
 
\section{Acknowledgements}

The authors would like to thank Science and Technology Facilities Physics for Fundamental Physics and Computing resources provided by STFC's Scientific Computing Department, the European Research Council (ERC − 2010 − AdG Grant 267841) and FCT (Portugal) grants SFRH/BD/75558/2010 for support. We acknowledge PRACE for awarding access to the supercomputing resources SuperMUC and JUQUEEN based in Germany.

\bibliography{Moon_MiniMag_Simulations_Bibliography}

\end{document}